\documentclass{article}
\usepackage{graphicx} 
\usepackage{subcaption} 
\usepackage{xcolor} 
\usepackage{amsmath} 
\usepackage{amssymb}

\usepackage{authblk}
\usepackage{multirow}

\usepackage[utf8]{inputenc}
\usepackage[numbers]{natbib}  
\usepackage[a4paper, top=0.8in, bottom=0.8in, left=0.8in, right=0.8in]{geometry}

\usepackage{amsmath}
\renewcommand{\eqref}[1]{ (\ref{#1})}

\title{Higher order Jacobi method for solving system of linear equations}
\author[1,2]{Nithin Kumar Goona}
\author[3,4]{Lama Tarsissi}
\affil[1]{College of Engineering, University of Texas at El Paso, USA}
\affil[2]{Adbutvaahak Pvt. Ltd., Bangalore, India}
\affil[3]{Dept.Sciences and Engineering-Sorbonne University Abu Dhabi, United Arab Emirates}
\affil[4]{LIGM, Univ Gustave Eiffel, CNRS, ESIEE Paris, F-77454 Marne-la-Vallée, France}

\date{}

\begin{document}
\maketitle

\section*{Abstract}

This work proposes a higher-order iterative framework for solving matrix equations, inspired by the structure and functionality of neural networks. A modification of the classical Jacobi iterative method is introduced to compute higher-order coefficient matrices through matrix-matrix multiplications. The resulting method, termed the higher order Jacobi method (HOJM), structurally resembles a shallow linear network and allows direct computation of the inverse of the coefficient matrix. Building on this, an iterative scheme is developed that allows efficient resolution of system variations without recomputing the coefficients, once the network parameters are trained for a known system. This iterative process naturally assumes the form of a deep recurrent neural network. The proposed approach goes beyond conventional physics-informed neural networks (PINNs) by providing an explicit, training-free definition of network parameters rooted in physical and mathematical formulations. Computational analysis on GPU reveals significant enhancement in the order of complexity, highlighting a compelling and transformative direction for advancing algorithmic efficiency in solving linear systems. This methodology opens avenues for interpretable and scalable solutions to physically motivated problems in computational science.

\section{Introduction}

Many real-world phenomena, ranging from fluid dynamics and heat transfer to financial modeling, electromagnetism, and more, are governed by Partial Differential Equations (PDEs). These equations describe how physical quantities evolve over space and time. However, solving PDEs analytically is often impractical, necessitating numerical methods to calculate their solutions.

To enable numerical solutions, partial differential equations (PDEs) \cite{smith1985numerical, ames2014numerical, johnson2009numerical} are discretized, converting the continuous problem into a system of algebraic equations through finite-difference or finite-element approximations. This discretization process often leads to a simple matrix equation of the form shown in Eq.\eqref{eq:matrix_equation}.

\begin{equation}
    A \cdot x = b
    \label{eq:matrix_equation}
\end{equation}

where $A$ is a system matrix derived from discretization, $x$ is the vector of unknowns, and $b$ represents known constraints or forcing terms. Efficiently solving this system is fundamental in computational science \cite{penrose1956best, prakash2003characteristic}, enabling simulations \cite{law2000simulation, bratley2011guide, king2025beyond} and optimizations \cite{dennis1996numerical, zainali2025modelling} in various engineering and scientific domains.

Although the discretization of PDEs can also lead to eigenvalue equations, this article will focus exclusively on a simpler matrix equation. There are several approaches to solving the matrix equation. Direct methods \cite{osterby1983direct, wilkinson1961error, gupta2002recent}, such as Gaussian elimination \cite{grcar2011mathematicians} and LU decomposition \cite{bartels1969simplex}, solve for $x$ exactly when feasible. Iterative methods \cite{varga1962iterative, hageman2014applied}, such as Jacobi \cite{varga1965iterative, tian2017jacobi} and conjugate gradient methods \cite{concus1976generalized, nocedal2006conjugate, nazareth2009conjugate}, are effective for large sparse systems. The AI driven approach, physics-informed neural networks (PINNs) \cite{raissi2019physics, cuomo2022scientific, cai2021physics}, leverages deep learning to approximate solutions while respecting the underlying physical laws.

This article presents a variation of the Jacobi method that can harness emerging hardware, such as GPUs being developed for AI to solve matrix equations. With this approach, higher-order system matrices are computed recursively, ultimately forming a deep linear network. The weights and biases are explicitly derived, offering deeper insight into PINNs, which are often treated as black boxes. It also eliminates the significant time and computational cost required for training. Notably, this higher-order method (HOM) can also be seamlessly extended to handle dense matrices, a scenario where traditional Jacobi methods are typically less effective.

Although the current state of physics-informed neural networks largely emphasizes physics-informed training \cite{karniadakis2021physics, nabian2021efficient}, our work shifts focus to intuitive, physics-informed explicit generation of weights and biases, paving the way for a more interpretable framework where the physics itself defines the structure of the network.

This study demonstrates explicit weight and bias construction for linear systems. Future extensions to nonlinear systems, building upon the linear case as a foundation, can offer even deeper insights into PINNs, particularly where the governing physics defines the nonlinear transformation within the network.

\section{The Higher Order Method}

The Jacobi approach to solving a matrix equation follows a decomposition of the matrix $A$ into its diagonal $D$, lower $L$ and upper $U$ matrices. The solution is then iteratively computed as shown in Eq.\eqref{eq:jacobi_iteration}, until the solution converges within a desirable tolerance.

\begin{equation}
    x^{(k+1)} = D^{-1} \cdot ( b - (L+U) \cdot x^{(k)} )
    \label{eq:jacobi_iteration}
\end{equation}

where $k$ and $k+1$ represent the current and next solution guesses respectively in the iterative process for $k \geq 0$.

\subsection{The higher order Jacobi method for solving matrix equation}

To facilitate the application of the higher-order version of the Jacobi iteration method, the system matrix $A$ and the source vector $b$ are normalized with respect to the diagonal matrix $D$ as defined in Eq.\eqref{eq:jacobi_iteration}. The resulting reformulated system is presented in Eq.\eqref{eq:jacobi_reformed}.

\begin{equation}
    x^{(k+1)} = -(A_n-I) \cdot x^{(k)} + b_n
    \label{eq:jacobi_reformed}
\end{equation}

where $I$ is the identity matrix. $A_n$ and $b_n$ are normalized matrices of $A$ and $b$, respectively.

Further simplification of Eq.\eqref{eq:jacobi_reformed} to express the next guess $x^{(k+1)}$ as a linear combination of the current guess $x^{(k)}$ leads to Eq.\eqref{eq:jacobi_simplified}.

\begin{equation}
    x^{(k+1)} = \alpha \cdot x^{(k)} + \beta
    \label{eq:jacobi_simplified}
\end{equation}

where

\begin{equation}
    \alpha = -(A_n-I)
    \label{eq:alpha}
\end{equation}

and

\begin{equation}
    \beta = b_n
    \label{eq:beta}
\end{equation}

This transformation ensures structured and scalable computation of the higher order coefficient matrices.

Starting with an initial guess $x^{(0)}$, the solution after the first iteration $x^{(1)}$ is calculated as shown in Eq.\eqref{eq:jacobi_simplified_first_guess}.

\begin{equation}
    x^{(1)} = \alpha \cdot x^{(0)} + \beta
    \label{eq:jacobi_simplified_first_guess}
\end{equation}

Similarly, the solution in second iteration $x^{(2)}$ is computed using the solution from the first iteration $x^{(1)}$ as shown in Eq.\eqref{eq:jacobi_simplified_second_guess}.

\begin{equation}
    x^{(2)} = \alpha \cdot x^{(1)} + \beta
    \label{eq:jacobi_simplified_second_guess}
\end{equation}

However, substituting Eq.\eqref{eq:jacobi_simplified_first_guess} into Eq.\eqref{eq:jacobi_simplified_second_guess}, corresponding to the second-order approximation, enables direct computation of the second iterate $x^{(2)}$ from the initial guess $x^{(0)}$, resulting in the following:

\begin{equation}
    x^{(2)} = \alpha \cdot (\alpha \cdot x^{(0)} + \beta)+ \beta
    \label{eq:jacobi_simplified_second_guess_from_initial_guess}
\end{equation}

Expanding Eq.\eqref{eq:jacobi_simplified_second_guess_from_initial_guess} results in Eq.\eqref{eq:jacobi_simplified_second_guess_reformed},

\begin{equation}
    x^{(2)} = \alpha^2 \cdot x^{(0)} + (\alpha + I) \cdot \beta
    \label{eq:jacobi_simplified_second_guess_reformed}
\end{equation}

Simplification of Eq.\eqref{eq:jacobi_simplified_second_guess_reformed} yields Eq.\eqref{eq:jacobi_simplified_second_guess_simplified}, explicitly illustrating the linear dependency of the second iterate $x^{(2)}$ on the initial guess $x^{(0)}$.

\begin{equation}
    x^{(2)} = \alpha^{(1)} \cdot x^{(0)} + \beta^{(1)}
    \label{eq:jacobi_simplified_second_guess_simplified}
\end{equation}

where

\begin{equation}
    \alpha^{(1)} = \alpha^2
    \label{eq:alpha1}
\end{equation}

and

\begin{equation}
    \beta^{(1)} = (\alpha + I) \cdot \beta
    \label{eq:beta1}
\end{equation}

Further generalizing Eq.\eqref{eq:jacobi_simplified_second_guess_simplified}, results in Eq.\eqref{eq:first_higher_order}

\begin{equation}
    x^{(2k+2)} = \alpha^{(1)} \cdot x^{(2k)} + \beta^{(1)}
    \label{eq:first_higher_order}
\end{equation}

which allows iterations to be performed using the second-order coefficients, $\alpha^{(1)}$ and $\beta^{(1)}$, respectively, instead of the first-order coefficients $\alpha$ and $\beta$ for $k \geq 0$.
Consequently, a single iteration with the second-order coefficients is equivalent to two iterations with the first-order coefficients.

However, Eq.\eqref{eq:jacobi_simplified_second_guess_simplified} takes the same form as Eq.\eqref{eq:jacobi_simplified_first_guess}, that is, it has a linear dependence on the initial guess. Following the same procedure of calculating second higher-order coefficients, recursively substituting the higher order equation into itself, the general form of the higher order Jacobi method (HOJM) is presented in Eq.\eqref{eq:higher_order}.

\begin{equation}
    x^{(2^k)} = \alpha^{(k)} \cdot x^{(0)} + \beta^{(k)}
    \label{eq:higher_order}
\end{equation}

where the recursive definitions for the higher-order coefficients are given by

\begin{equation}
    \alpha^{(k)} = {(\alpha^{(k-1)})}^2
    \label{eq:alphak}
\end{equation}

and

\begin{equation}
    \beta^{(k)} = (\alpha^{(k-1)} + I) \cdot \beta^{(k-1)}
    \label{eq:betak}
\end{equation}

Consequently, the solution after $2^k$ iterations with first-order coefficients is equivalent to 1 iteration with the $k^{th}$-order coefficients.

Using Eq.\eqref{eq:higher_order}, any higher order guess $x^{(2^k)}$ can be calculated directly from the initial guess $x^{(0)}$. However, each higher-order guess $x^{(2^k)}$ can also be obtained from the previous higher-order guess $x^{(2^{k-1})}$ as shown in Eq.\eqref{eq:higher_order_update}, for faster convergence.

\begin{equation}
    x^{(2^k)} = \alpha^{(k)} \cdot x^{(2^{k-1})} + \beta^{(k)}
    \label{eq:higher_order_update}
\end{equation}

The exponential reduction in the number of iterations comes at the cost of an increase in the computational complexity of matrix-matrix multiplication. Each higher-order step requires multiplying two square matrices, whereas traditional Jacobi iterations only involve multiplying a square matrix with a column vector.

For large matrices, block matrix multiplication and GPU acceleration with parallel processing significantly reduce real-world execution time of computing the higher-order coefficients.

\subsection{Finding inverse of coefficient matrix using the higher order Jacobi method}

The expanded form of Eq.\eqref{eq:higher_order_update} exhibits a clear structural similarity to a deep linear network comprising $k$ layers, with analytically defined weights and biases at each layer. It is well-established that any linear network, irrespective of its depth, can be equivalently represented as a single-layer network. In the context of the proposed higher order Jacobi method, this single-layer equivalent corresponds to the inverse of the normalized coefficient matrix $A_n$, as elaborated in the following discussion.

In Eq.\eqref{eq:higher_order}, the solution uniquely depends on the bias term $\beta^{(k)}$, while the other term is arbitrary, as the initial guess can be arbitrary, even though $\alpha^{(k)}$ is uniquely defined. The term $\alpha$ times $x$ can become insignificant with a null vector as an initial guess. Thus, the multiplication term can be omitted, resulting in Eq.\eqref{eq:higher_order_truncated}.

\begin{equation}
    x^{(2^k)} = \beta^{(k)}
    \label{eq:higher_order_truncated}
\end{equation}

Expanding Eq.\eqref{eq:higher_order_truncated} recursively results in Eq.\eqref{eq:higher_order_truncated_expanded}.

\begin{equation}
    x^{(2^k)} = (\alpha^{(k-1)} + I) \cdot (\alpha^{(k-2)} + I) \quad \ldots \quad (\alpha^{(1)} + I) \cdot (\alpha + I) \cdot \beta^{(0)}
    \label{eq:higher_order_truncated_expanded}
\end{equation}

where

\begin{equation}
    \beta^{(0)} = \beta
    \label{eq:beta0}
\end{equation}

Expressing Eq.\eqref{eq:higher_order_truncated_expanded} using compact product notation leads to the form shown in Eq.\eqref{eq:higher_order_truncated_generalised}.

\begin{equation}
    x^{(2^k)} = \left( \prod_{i=k-1}^{0} (\alpha^{(i)} + I) \right) \cdot \beta
    \label{eq:higher_order_truncated_generalised}
\end{equation}

where

\begin{equation}
    \alpha^{(0)} = \alpha
    \label{eq:alpha0}
\end{equation}

Abstracting Eq.\eqref{eq:higher_order_truncated_generalised}, results in Eq.\eqref{eq:inverse_equation}.

\begin{equation}
    x^{(2^k)} = A_n^{-1} \cdot \beta
    \label{eq:inverse_equation}
\end{equation}

where 

\begin{equation}
    A_n^{-1} = \prod_{i=k-1}^{0} (\alpha^{(i)} + I)
    \label{eq:inverse}
\end{equation}

The explicit closed-form representation of the inverse of matrix $A_n$, as shown in Eq.\eqref{eq:inverse}, is derived from the Jacobi method, which proves the equivalence of direct and iterative methods of solving the matrix equation, when necessary conditions for convergence \cite{james1973convergence, costa2022convergence, tekriwal2023verified} are met in the regular Jacobi iteration method. In addition, the order of accuracy can also be selected; higher the order of coefficient matrices, higher the accuracy. Eq.\eqref{eq:inverse_equation} can be renormalized to obtain the actual inverse of the original system matrix $A$.

\subsection{The higher order Jacobi method as a deep network}

Although Eq.\eqref{eq:higher_order_update} takes the same form as a deep linear network with weights and biases, Eq.\eqref{eq:higher_order_truncated_generalised} represents its transformation into a single-layer linear network, especially one without biases.

For an arbitrary forcing matrix $b$, the higher-order method produces a solution the same as the regular Jacobi iteration method in the equivalent number of iterations.
However, in real-world applications, the system matrix $A$ is not constant and can change from problem to problem due to the material properties embedded in different PDEs \cite{nagel2014numerical} or due to mesh refinement \cite{ji2010new}, which requires recalculation of the higher-order matrices, which may be computationally intensive.

Even with single change in a value of the original coefficient matrix, involve recalculation of the higher-order coefficient matrices using matrix-matrix multiplication, making it inefficient for dynamically changing systems, especially for non-linear systems, where values of the coefficient matrix depend on the solution. At this stage, iterative methods constitute the only viable pathway for rigorously understanding the structural necessity and functional dynamics of layered architectures in neural networks. A representative process is presented in detail in the following section.

\subsection{Induced source iterative scheme as an explicit recurrent neural network}

The coefficient matrix $A$ may undergo significant changes in scenarios such as mesh refinement. However, the following scheme is designed for cases where the matrix varies gradually, as encountered in nonlinear systems. Instead of recalculating the modified higher-order coefficients, the goal is to adjust the source matrix \cite{goona2021distributed, goona2023distributed} so that solving a system with unmodified higher-order coefficients and modified source yields the same solution as solving the system with updated coefficients and the original source. This approach eliminates the need for recomputing higher-order coefficients while maintaining solution consistency.

Let the modified source $b_m$ consist of the original source $b$ and the induced source $b_i$ as defined in Eq.\eqref{eq:bm}.
\begin{equation}
    b_m = b + b_i
    \label{eq:bm}
\end{equation}

where $b$ is the original source matrix for which the new system is to be solved and $b_i$ is the induced source, initialized as a null matrix and computed iteratively.

This iterative approach introduces the following series of intermediate steps in which the solution is first obtained using the known inverse of the coefficient matrix, as shown in Eq.\eqref{eq:induced_source_iteration1}. The solution-dependent coefficient matrix $A_d$ is then calculated as shown in Eq.\eqref{eq:nonlinear_iteration}. Followed by one regular Jacobi iteration step with the new system matrix $A_d$ as shown in Eq.\eqref{eq:induced_source_iteration2}. Finally, the induced source $b_i$ is updated as shown in Eq.\eqref{eq:induced_source_iteration3}.

\begin{equation}
    x^{(k)} = A^{-1} \cdot (b + b_i^{(k)})
    \label{eq:induced_source_iteration1}
\end{equation}
\begin{equation}
    A_{d}^{(k)} = \digamma (x^{(k)})
    \label{eq:nonlinear_iteration}
\end{equation}
\begin{equation}
    x^{(k+1)} = -(A_{d}^{(k)}-I) \cdot x^{(k)} + b
    \label{eq:induced_source_iteration2}
\end{equation}
\begin{equation}
    b_i^{(k+2)} = \omega \cdot b_i^{(k)} + (1-\omega) \cdot (A-A_{d}^{(k)}) \cdot x^{(k+1)}
    \label{eq:induced_source_iteration3}
\end{equation}

where $A^{-1}$ is a known inverse, calculated using the higher order method or other techniques, $A_{d}$ is the coefficient matrix dependent on the solution for the new system, $\digamma$ is a physics defined function that reflects the dependency of the coefficient matrix on the solution, that is, the nonlinearity, and $\omega$ is a scalar relaxation factor that facilitates smoother convergence.

Eqs.\eqref{eq:induced_source_iteration1} -\eqref{eq:induced_source_iteration3} involve only matrix-vector multiplications, eliminating computationally expensive matrix-matrix multiplications. These equations are applied iteratively until convergence is achieved with a predefined tolerance, similar to a recurrent network structure where $\digamma$ plays the role of the nonlinear transformation.

\section{Results}

\begin{figure}[htbp]
    \centering
    \begin{subfigure}[b]{0.48\textwidth}
        \centering
        \includegraphics[width=\linewidth]{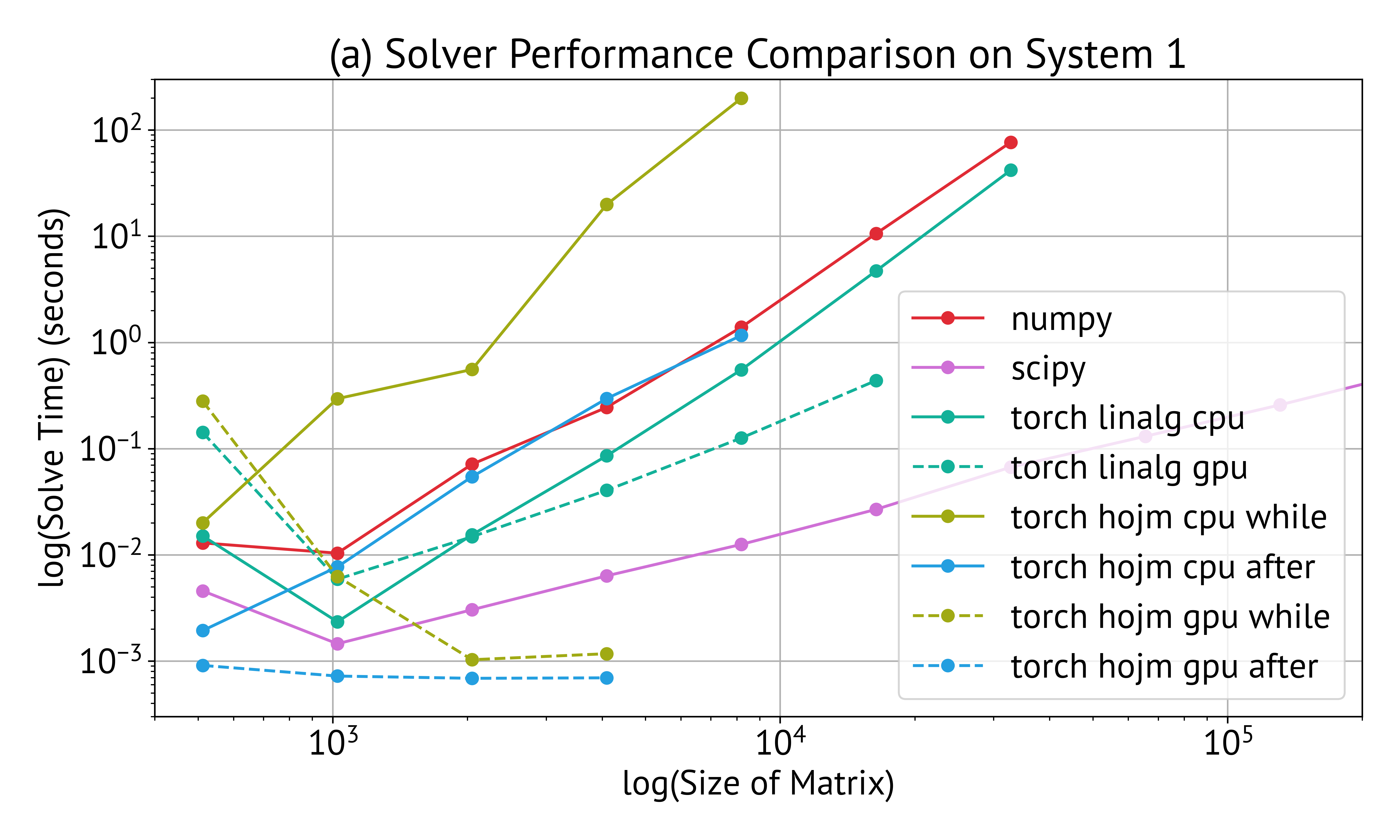}
    \end{subfigure}
    \hfill
    \begin{subfigure}[b]{0.48\textwidth}
        \centering
        \includegraphics[width=\linewidth]{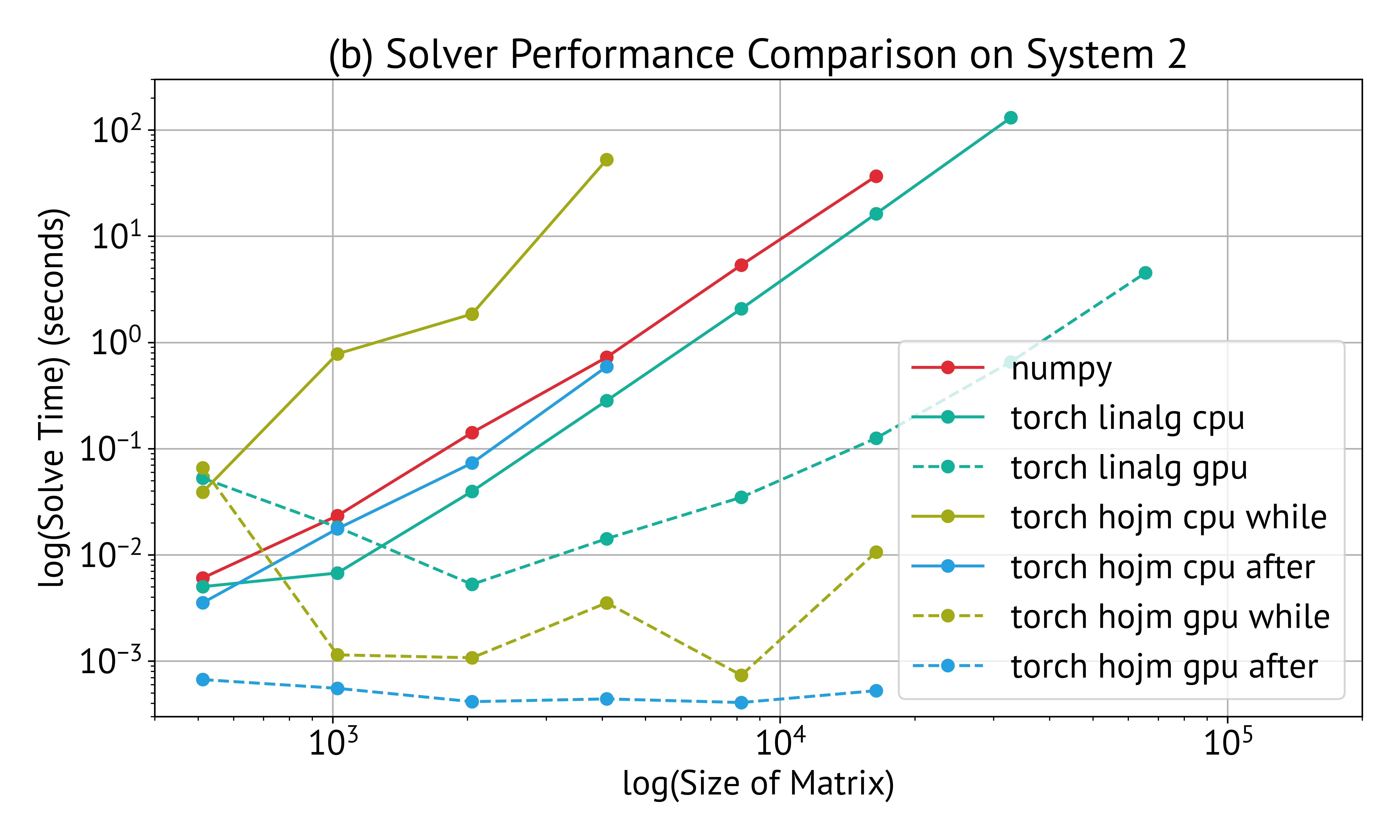}
    \end{subfigure}

    \vspace{1em}  

    \begin{subfigure}[b]{0.48\textwidth}
        \centering
        \includegraphics[width=\linewidth]{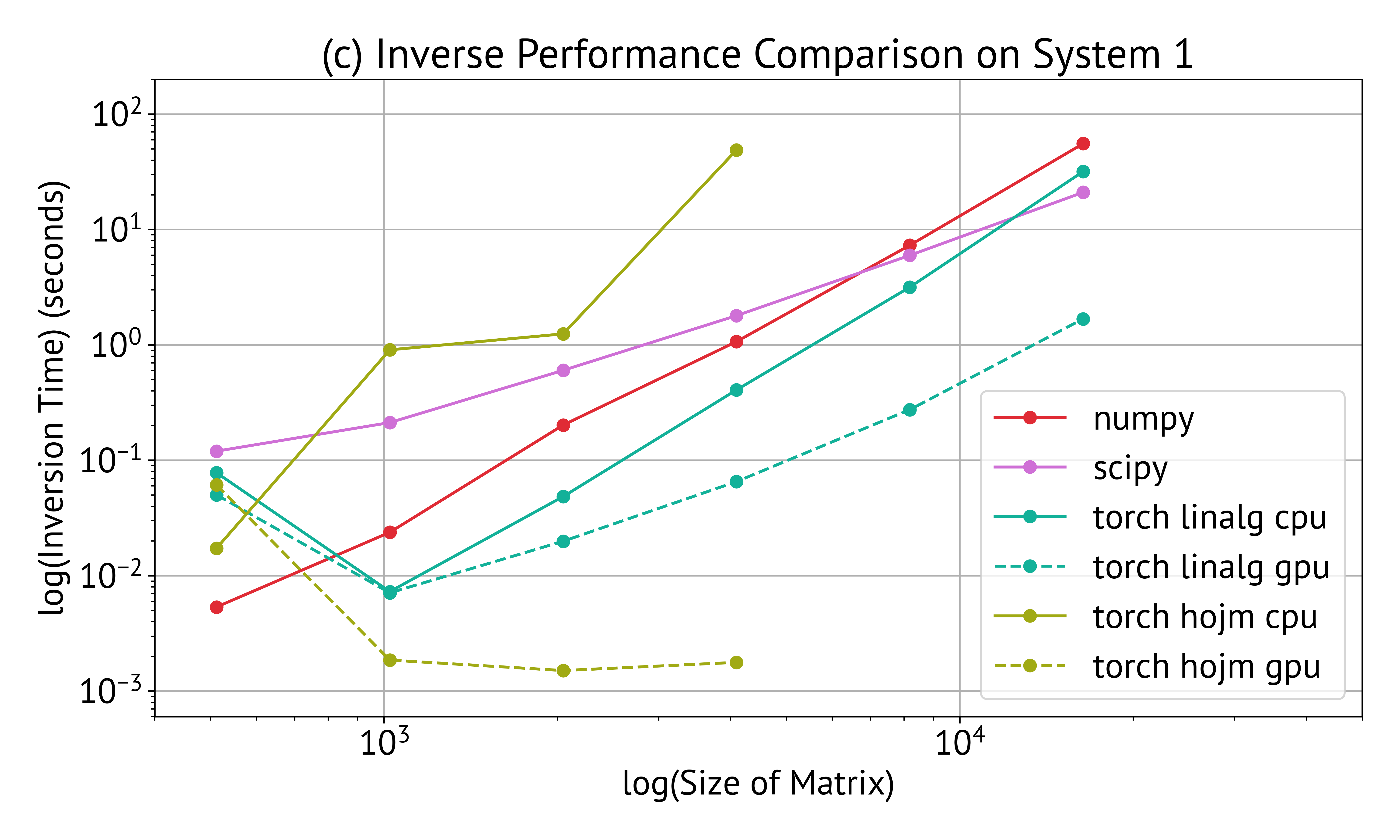}
    \end{subfigure}
    \hfill
    \begin{subfigure}[b]{0.48\textwidth}
        \centering
        \includegraphics[width=\linewidth]{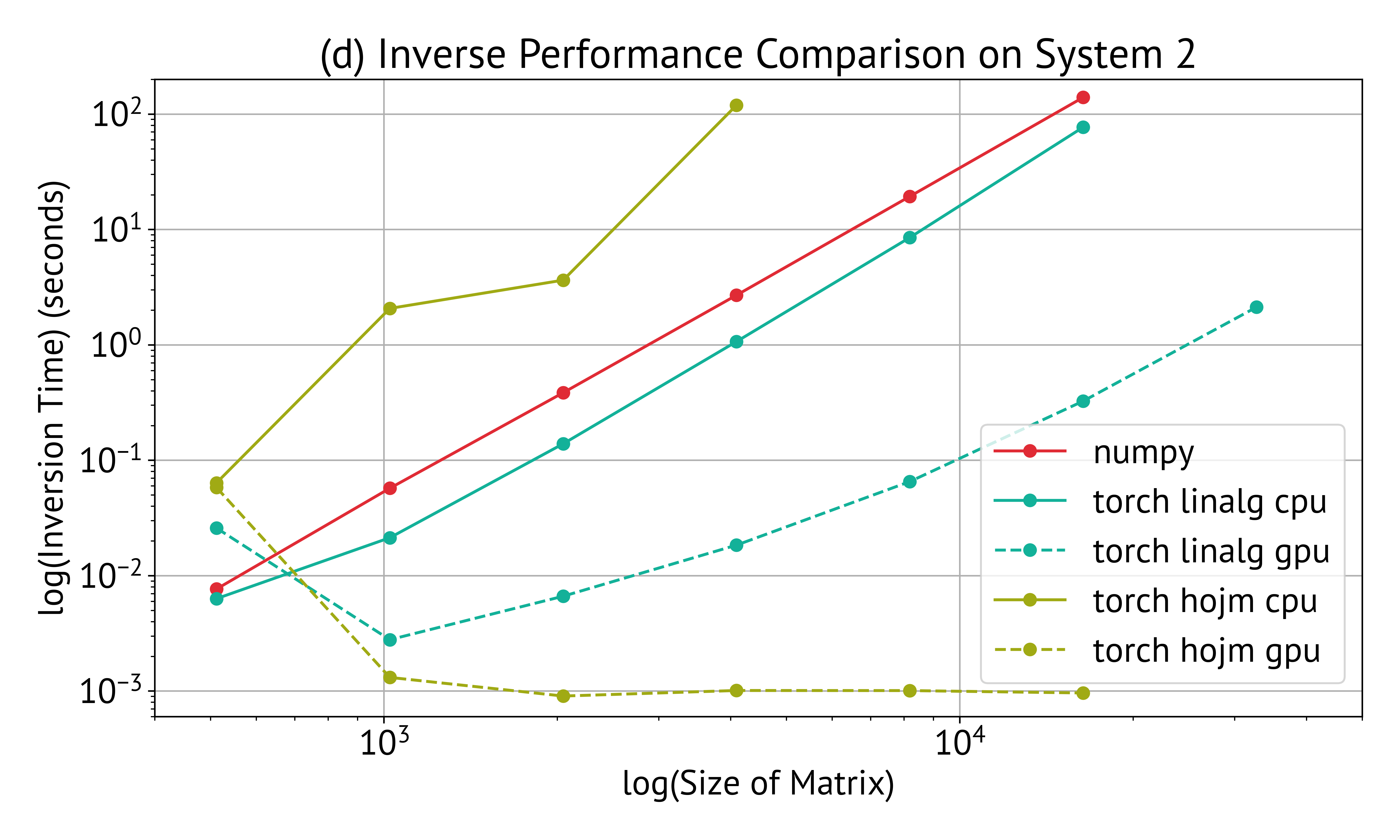}
    \end{subfigure}

    \caption{Performance Comparision of built-in methods and the proposed higher order Jacobi method in solving system of linear equations and finding matrix inverse, (a) Solving matrix equation on System 1, (b) Solving matrix equation on System 2,
    (c) Finding matrix inverse on System 1, and (d) Finding matrix inverse on System 2.}
    \label{fig:performance}
\end{figure}

A comparison of the time taken to solve a matrix equation and to invert a matrix with increasing matrix size using different methods is shown in Fig. \ref{fig:performance}. The results in Fig. \ref{fig:performance}(a) \& \ref{fig:performance}(c) were obtained on a computer system (System 1) with lower specifications, while those in Fig. \ref{fig:performance}(b) \& \ref{fig:performance}(d) were obtained on a higher-end system (System 2). The configurations of both systems are presented in Table \ref{tab:config}. The matrix equation is solved using the built-in solve method from the linear algebra subpackage of each Python library for the existing methods, and using the built-in matrix-matrix multiplication functionality in PyTorch for the higher order Jacobi method implementation. The matrix inverse is similarly computed using the respective built-in inverse functions for existing methods, while the higher order Jacobi method approximates the inverse through iterative matrix-matrix operations. The reported inverse times also account for the re-normalization of the coefficient product. The x-axis represents the matrix size, while the y-axis shows the time taken to compute the solution. The results are presented in log-log scale. All the calculations are performed with 32-bit single presision data formats.

The horizontal extent of each curve depends on the available system memory for the corresponding operation, whereas the vertical range is capped to 20 seconds for any single operation whether it is solving a single matrix equation or inverting a matrix. The coeffecient matrix $A$ is constructed to be sparse. Irrespective of the size of the matrix, 20 higher coeffecients are calculated while producing the higher order Jacobi results. 20 higher order iterations is equivalent to 1 million regular Jacobi iterations. The accuracy in the solution from 20 higher order iterations is equivalent to accuracy of Jacobi method after million iterations.

The CPU implementations of the methods are shown using solid lines, while the GPU implementations are represented by dotted lines. In System 1, a Conda environment was created, and the required Python packages, as listed in Table \ref{tab:config}, were installed. In contrast, System 2 utilizes a default docker pytorch container, as it is part of a high-performance GPU cluster. It is due to this reason the package versions are slightly different and scipy is not available on System 2. Since System 2 is primarily designed for GPU computing, a dedicated GPU is allocated when the code is executed, whereas its CPU may have been reserved for other tasks.

The runtime empirical computational complexity for each curve is calculated using the last two data points. If a sudden jump is observed in the last 3 data points, the middle point is excluded for the calculation. The corresponding results are presented in Table \ref{tab:results}. The best order in each row has be highlighted.

From Fig. \ref{fig:performance} and Table \ref{tab:results}, it is evident that the higher order Jacobi implementation using PyTorch GPU tensors performs the best, regardless of the system or whether a solve or inverse method is used. This performance advantage is attributed to the higher order Jacobi method’s ability to leverage the parallelism inherent in matrix-matrix multiplications.

\begin{table}[htbp]
\centering
\caption{Configuration of the two compute systems used to compare the performance.}
\begin{tabular}{ccc}
\hline
 & System 1 & System 2 \\
\hline
CPU & Intel Core i7-10870H & Intel Xeon Sapphire Rapids \\
Cores & 8 & 2 $\times$ 56 \\
System RAM & 32 GB & 2 TB \\
GPU & \begin{tabular}[c]{@{}c@{}}NVIDIA GeForce RTX 3070 \\ Laptop Version\end{tabular} & \begin{tabular}[c]{@{}c@{}}NVIDIA Hopper H100 \\ Tensor Core\end{tabular} \\
Graphics Memory & 8 GB & 80 GB \\
Operating System & Ubuntu 24.04.1 LTS & \begin{tabular}[c]{@{}c@{}}Red Hat Enterprise \\ Linux (RHEL) 9.2\end{tabular} \\
Python Package Versions & \begin{tabular}[c]{@{}c@{}}python = 3.10.18, numpy = 2.2.6, \\ pytorch = 2.5.1, pytorch-cuda = 11.8, \\ scipy = 1.15.2\end{tabular} & \begin{tabular}[c]{@{}c@{}}python = 3.10.13, numpy = 1.26.3, \\ pytorch = 2.2.1, pytorch-cuda = 12.1\end{tabular} \\
\hline
\end{tabular}
\label{tab:config}
\end{table}

\begin{table}[htbp]
\centering
\caption{Performance of solving matrix equation and finding inverse using built-in methods of Numpy, Scipy, PyTorch CPU/GPU and the higher order Jacobi method using PyTorch built-in matrix-matrix multiplication.}
\begin{tabular}{cllllll}
\hline
\multicolumn{1}{l}{}                                                               & \multicolumn{1}{c}{\begin{tabular}[c]{@{}c@{}}NumPy\\ linalg\end{tabular}} & \multicolumn{1}{c}{\begin{tabular}[c]{@{}c@{}}Scipy\\ linalg\end{tabular}} & \multicolumn{2}{c}{\begin{tabular}[c]{@{}c@{}}Pytorch\\ linalg\end{tabular}} & \multicolumn{2}{c}{\begin{tabular}[c]{@{}c@{}}Pytorch\\ HOJM\end{tabular}}                            \\ \hline
\multirow{2}{*}{\begin{tabular}[c]{@{}c@{}}Solving Matrix\\ Equation\end{tabular}} &  &   & \multicolumn{1}{c}{CPU}  & \multicolumn{1}{c}{GPU}  & \multicolumn{1}{c}{CPU} & \multicolumn{1}{c}{GPU} \\
&   &   &   &  & \multicolumn{2}{c}{while}  \\

System 1 & $ 9.5e^{-12}n^{2.86} $ & $ 3.4e^{-08}n^{1.28} $ & $ 2.5e^{-13}n^{3.15} $ & $ 1.2e^{-08}n^{1.80} $ & $ 2.0e^{-11}n^{3.32} $ & \textbf{\boldmath$2.4e^{-04}n^{0.19}$} \\
System 2 & $7.2e^{-11}n^{2.78}$ & \multicolumn{1}{c}{--} & $3.9e^{-12}n^{3.00}$ & $1.8e^{-13}n^{2.78}$ & $5.6e^{-10}n^{3.04}$ & \textbf{\boldmath$4.8e^{-06}n^{0.79} $}\\
\multirow{1}{*} &   &   &   &  & \multicolumn{2}{c}{after}  \\
System 1 &  &  &  &  & $ 2.2e^{-08}n^{1.98} $  & \textbf{\boldmath$6.1e^{-04}n^{0.01}$} \\
System 2 &  &  &  &  & $7.5e^{-12}n^{3.02}$ & \textbf{\boldmath$1.4e^{-05}n^{0.37}$} \\\hline

\begin{tabular}[c]{@{}c@{}}Finding Matrix\\ Inverse\end{tabular} &  &  & CPU & GPU & \multicolumn{1}{c}{CPU} & \multicolumn{1}{c}{GPU} \\
System 1 & $2.6e^{-11}n^{2.93}$ & $4.7e^{-07}n^{1.81}$ & $3.0e^{-13}n^{3.33}$ & $1.6e^{-11}n^{2.61}$ & $2.0e^{-09}n^{2.88}$ & \textbf{\boldmath$2.5e^{-04}n^{0.24}$} \\
System 2 & $1.3e^{-10}n^{2.85}$ & \multicolumn{1}{c}{--} & $3.1e^{-12}n^{3.18}$ & $1.3e^{-12}n^{2.71}$ & $3.2e^{-09}n^{2.93}$ & \textbf{\boldmath$1.8e^{-03}n^{-0.07}$} \\ \hline
\end{tabular}
\label{tab:results}
\end{table}

\section{Discussions}

The computational complexity of solving matrix equations depends primarily on two factors \cite{buss1999computational, golub2013matrix, Papad2003Comp}, the sparsity of the coefficient matrices and the type of method used, direct or iterative. The system matrices built using PDEs are usually sparse with very few nonzero elements in each row and are diagonally dominant. The computational complexity of solving such sparse systems is much lower when compared with their denser counterparts, such as Hamiltonian matrices in computational quantum chemistry. This is due to the reduced number of calculations performed in each row. A comprehensive comparison of different computational complexities is shown in Table \ref{tab:cc}.

\begin{table}[ht]
    \centering
    \caption{Comparison of computational complexities of solving matrix equations and matrix multiplications using different known methods.}
    \begin{tabular}{ll}
    \hline
    \multicolumn{1}{c}{Method} & \multicolumn{1}{c}{Complexity} \\
    \hline
    Solving Matrix Equation & \\
    \quad Gaussian Elimination (Dense) & $O(n^3)$ \\
    \quad LU Decomposition (Dense) & $O(n^3)$ \\
    \quad Jacobi Iteration (Dense, $k \sim n$) & $O(n^3)$ \\
    \quad Jacobi Iteration (Sparse, $s \ll n$, $k \ll n$) \cite{song2016joint} & $O(kns)$  \\
    \quad Conjugate Gradient (SPD matrix) \cite{chandra1978conjugate, nazareth2009conjugate} & $O(n^2)$ to $O(n \log n)$ \\
    \quad Multigrid Method \cite{trottenberg2001multigrid} & $O(n)$ \\
    Matrix-Matrix Multiplication & \\
    \quad Naive Matrix Multiplication & $O(n^3)$ \\
    \quad Strassen's Algorithm \cite{huss1996implementation} & $O(n^{\log_2(7)}) \approx O(n^{2.81})$ \\
    \quad Google's AlphaEvolve \cite{novikov2025alphaevolve} & $O(n^{\log_4(48)}) \approx O(n^{2.79})$ \\
    \quad Coppersmith-Winograd Algorithm \cite{coppersmith1982asymptotic} & $O(n^{2.376})$ \\
    \quad Improved Variants of Coppersmith-Winograd \cite{williams2012multiplying} & $O(n^{2.3728596})$ \\
    Matrix-Vector Multiplication & $O(n^2)$\\
    \hline
    \end{tabular}
    \label{tab:cc}
\end{table}

The proposed higher-order method has the same computational complexity as matrix-matrix multiplication. Although its naive approach is an order of magnitude higher than matrix-vector multiplication, matrix-matrix multiplication is a highly parallelizable operation \cite{agarwal1994high, choi1998new, osama2023stream, riz2022opti}.

Moreover, the theoretical computational complexity of multiplying large matrices has improved significantly over the years, falling from the default $O(n^{3})$ to the current known limit of $O(n^{2.376})$ \cite{coppersmith1982asymptotic} for large matrices. However, practical and scalable implementations even with the help of deep learning algorithms \cite{novikov2025alphaevolve,kauers2025consequences}still tend to perform more closely to the Strassen algorithm in practice.

However, the higher order Jacobi method takes advantage of the parallelizability of matrix-matrix multiplication, unlike the current sequential approaches, and therefore demonstrates exceptional performance when implemented on GPUs.


As coefficient matrices become denser with recursive squaring, pruning techniques \cite{JMLR:v22:21-0366, 10643325} used in traditional neural networks can be adopted to make dense higher-order matrices much more compact. However, with recent developments in parallel computing and the ever-increasing affordability of GPUs \cite{10.1145/3503221.3508431, 10938448} and even photonic circuits \cite{Huang:23}, matrix-matrix multiplications can be further accelerated.

The proposed higher-order method for matrix inversion structurally aligns with a deep linear network. Once the coefficients are explicitly derived for a known system, the method enables efficient resolution of variations of that system without the need for computationally intensive re-computation of the coefficients. The recurrent iterative scheme inherently adopts the form of a deep recurrent neural network. This framework aims to provide analytical insight into the explicit definition of network parameters in physics informed neural networks, thereby circumventing the training process and significantly reducing computational overhead.

\section{Conclusion}

Physical systems are fundamentally governed by differential equations, and recent advances have shown growing interest in Physics-Informed Neural Networks (PINNs) for solving such equations. However, in most existing frameworks, the term ``physics-informed'' relates primarily to the incorporation of physical laws into the training objective, rather than to the explicit definition of network parameters such as weights, biases, and activation functions.

This work presents a direction toward interpreting and defining the inner workings of neural networks through an explicit analytical formulation. In particular, a modified Jacobi iteration method is developed to compute higher-order coefficient matrices using matrix-matrix multiplications. The resulting formulation, referred to as the higher order Jacobi method, exhibits the structural characteristics of a shallow linear network with a limited number of layers. Through this formulation, it is also possible to compute the inverse of the original coefficient matrix directly via matrix operations, eliminating the need for iterative inversion procedures.

In addition, an iterative scheme is proposed in which the network parameters, once determined for a known system, can be reused to solve variations of the system without recomputing the coefficients. This capability is especially advantageous in applications where the underlying system undergoes parametric changes. The iterative process takes the form of a deep recurrent neural network, offering a principled and efficient approach that maintains consistency across related systems.

Computational analysis indicates that the proposed methods lead to exceptional improvement in the order of complexity when implemented on GPU. This signifies a significant algorithmic advancement in solving matrix equations, with reduced computational cost and improved scalability. The presented methodology not only connects traditional numerical methods with neural architectures, but also contributes to interpretable \cite{rudin2022interpretable, esterhuizen2022interpretable} and training-free network designs grounded in physical and mathematical principles.

Future directions include the extension of this framework to nonlinear systems and time-dependent differential equations, and eigenvalue problems, enabling broader applications in scientific computing and model-driven learning.

\section*{Code Availability}

A simple implementation of the proposed method and the code used to generate the data and the plots are available at the following link: https://github.com/nithingoona/HOJM

\bibliography{main}

\end{document}